\documentclass[12pt,prb]{revtex4}

\usepackage[english]{babel}
\usepackage[intlimits]{amsmath}
\usepackage[dvips]{graphicx}

\begin{document}

\begin{titlepage}

\title{ Lattice Models of Ionic Systems with Charge Asymmetry}

\author{Maxim N. Artyomov}
\affiliation{Moscow State University, Moscow 119899, Russia}
\author{Vladimir Kobelev, Anatoly B. Kolomeisky}
\affiliation{Department of Chemistry, Rice University, Houston, Texas 77005}

\begin{abstract}

\noindent
The thermodynamics of a charge-asymmetric lattice gas of positive ions carrying  charge $q$ and negative ions with charge $-zq$  is investigated  using  Debye-H\"uckel theory. Explicit analytic and numerical calculations, which take into account  the formation of neutral and charged clusters  and cluster  solvation by the residual ions, are performed for $z=2$, 3 and 4. As charge asymmetry increases, the predicted critical point shifts  to lower temperatures and higher densities. This trend  agrees well with the results from recent Monte Carlo simulations for continuum  charge-asymmetric hard-sphere ionic fluids and with the corresponding  predictions from continuum Debye-H\"uckel theory. 

\end{abstract}

\maketitle

\end{titlepage}

\section{Introduction}

The nature of  critical phenomena in ionic systems has been a subject of numerous theoretical studies in recent years.\cite{fisher93,stell,levin96,weingartner01}  Due to the long-range nature of Coulombic interactions, construction of a full renormalization group treatment, which was so successful in describing critical behavior of non-ionic fluids, meets both conceptual and technical difficulties.\cite{fisher93} However, in recent years some  progress has been achieved in obtaining  physically reasonable,  well-based mean-field theories for ionic systems.\cite{fisher93,stell,levin96,weingartner01,levin02} These theoretical studies have been supported and, sometimes in substantial part, initiated by intensive Monte Carlo simulations \cite{panagiotopoulos, orkoulas99, camp99,lujiten01, pana02,yan02} of charged systems.

To investigate the thermodynamics of  ionic fluids, two main mean-field approaches have been developed. The first one\cite{fisher93,levin96} extends the pioneering work of Debye and H\"uckel\cite{debye} (DH)  on dilute  solutions   of strong electrolytes, while  the second approach\cite{stell} relies  on  integral equations for correlation functions. Analysis of thermodynamic energy bounds and comparison with the best Monte Carlo estimates for the critical parameters suggests that  the DH-based theory gives a better description of the thermodynamics of electrolytes, at least  in the critical region.\cite{levin96,zuckerman97}

The simplest model of ionic fluids, the so-called restrictive primitive model (RPM), considers a system of spherical equisized charged particles, half of them  carrying a  charge $q$ and the other half with  charge $-q$. The charge symmetry of this model plays a crucial role in the determination of its universality class and in the ability to obtain analytic solutions. This raises the   question of  how the breaking of the symmetry will affect the thermodynamics and critical properties of electrolyte systems. An important extension of the RPM is the charge-asymmetric primitive model, where the sizes of negative and positive particles are the same  while  absolute values of charges  for positive and negative ions are different. Recent   Monte Carlo simulations\cite{pana02,camp99} have revealed that, as charge asymmetry increases, the critical temperature $T_{c}$ of the  gas-liquid transition decreases, while the critical density $\rho_{c}$ grows.  However, most of the current theories give different predictions. Simple DH theory and the mean-spherical approximation (MSA) both predict no dependence on the asymmetry parameter $z$.\cite{levin96,MSA} In symmetric Poisson-Boltzmann and modified Poisson-Boltzmann  integral equation  methods\cite{sabir} the charge asymmetry hardly  changes  $T_{c}$ , while  $\rho_{c}$  increases. However, the absolute values of the critical parameters are very different from Monte Carlo   estimates. A field theoretical approach by Netz and Orland\cite{netz} predicts large increases in critical temperatures and, similarly, a large decrease in critical densities, in  sharp contrast with computer simulations trends. The only theory  that produces  reasonable  results for  the effect of charge asymmetry on thermodynamics and criticality of ionic systems, as judged by comparison with Monte Carlo computer simulations, is the DH approach  augmented by Bjerrum cluster formation and cluster-ion interactions (DHBjCI).\cite{banerjee02}

Lattice models, such as the Ising model, have played an important role in understanding  criticality in non-ionic systems. In recent years, lattice models have also attracted the attention of researchers as a tool for  investigating  thermodynamics and criticality in Coulomb systems.\cite{fisher93,dickman99,ciach01,kobelev02,brognara02,anisotropy} A systematic study of electrolytes on lattices, which utilizes the  Debye-H\"uckel approach, has been presented recently.\cite{kobelev02} In this work  the thermodynamics of a  $d$-dimensional system of equal numbers of positive and negative ions, i.e., a charge-symmetric lattice RPM, has been investigated. Specific calculations for Coulomb systems on three-dimensional lattices, which included ion pairing and ion-dipole interactions, predicted a gas-liquid phase separation at low densities. However, taking into account the lattice symmetry yielded a different scenario --- the phase diagrams of  electrolytes on simple cubic and body-centered cubic lattices show order-disorder phase transitions with a tricritical point, while  gas-liquid phase separation is suppressed. The introduction of charge asymmetry in lattice models of ionic fluids tends to suppress  the possibility of order-disorder phase transitions, and  gas-liquid phase coexistence  reappears, although  the position of critical point may  change.

In this paper, we present a thermodynamic  investigation of charge-asymmetric  lattice models  of electrolytes. By explicitly including  the  clustering of oppositely charged particles and ion-cluster interactions, we obtain phase diagrams for 2:1 and 3:1 lattice electrolytes, and we locate  the critical point for the  4:1 ionic system. Our results accord well with the trend obtained in recent Monte Carlo simulations and the continuum DHBjCI theory.\cite{banerjee02} The paper is organized as follows. In Sec.II we present an overview of our  thermodynamic approach to multicomponent charged species mixtures and we outline  the pure Debye-H\"uckel theory. The full theory, which accounts for charged and neutral cluster formation and their interactions  with the residual ions, for a $2:1$ system is presented  in detail in Sec.III. Sec.IV describes the general scheme of thermodynamic calculations for $3:1$ and $4:1$ lattice electrolytes. Finally,  the  results are discussed in Sec.V and our  conclusions are given in Sec.VI.

\section{Lattice Debye-H\"{u}ckel Theory of Charge-asymmetric Electrolytes}

\subsection{Thermodynamic Overview}

Consider a system of charged particles  on a three-dimensional  simple cubic lattice with a unit cell length  $a$, which  initially has  $N_0$ ions carrying a charge $-zq$ and $z N_0$ ions with a charge $q$, i.e., the total number of ions is $N=(z+1)N_{0}$. Because of the electrostatic interactions,  ions with opposite charges  tend to form clusters. As a result,  there will be many  species present in the system: dimers, trimers, etc., with respective charges $(-z+1)q, (-z+2)q,..,0$.  If  the number of particles of type $i$ is given by $N_{i}$, then we  define  $\rho_i=N_i/V$ and $\rho_i^*=\rho_i {\rm v}_0$ to be the number density and the reduced density of the $i$-th species, with ${\rm v}_0=a^3$ being the unit lattice cell volume. 

The Helmholtz free energy is central for the determination of the thermodynamic behavior of charge-asymmetric lattice electrolytes. It can be approximated  by summing consecutively the free energies describing the interactions between different species,
\begin{equation}
     F =  F^{Id} + \sum_i F_i ,
\end{equation}
where $ F^{Id}$ is the ideal lattice gas (entropic) term and $F_i$ is the electrostatic energy of the $i$-th species. Once the reduced free energy density $\bar f \equiv - F/k_BTV$ is known, the reduced chemical potentials for every component  $\bar\mu_i\equiv\mu/k_B T$ can be computed via
\begin{equation}\label{chem.potential}
\bar \mu_{i} = - \partial \bar f /\partial \rho_{i}.
\end{equation}
Finally, the reduced pressure is given by 
\begin{equation} \label{pressure}
\bar p \equiv p/k_BT=\bar f +\sum_i \rho_i \bar\mu_i. 
\end{equation}
Then the possible phase equilibria are  defined by matching pressures and chemical potentials for each component in different phases.

In multicomponent systems with charged particles  it is the electrochemical potential that must be equal in  coexisting vapor ($v$) and liquid ($l$) phases,\cite{banerjee02}  namely,
\begin{equation}
    \mu_{i,v}+q_i\phi_v = \mu_{i,l}+q_i\phi_l ,
\end{equation}
where $\phi_{v(l)}$ is the electrostatic potential in the corresponding phase, where, in  general, there is a  nonzero Galvani potential difference $\Delta \phi = \phi_v- \phi_l$ between the phases.\cite{galvani} However, for calculating phase equilibria in multicomponent systems of charged particles, it is more convenient to use the single-component thermodynamic picture.\cite{banerjee02} Since every thermodynamic phase is electroneutral, the multicomponent system with $N=(z+1)N_{0}$ ions can be viewed as a  single-component system of $N_{0}$ molecules, each of them consisting of one negative ion and $z$ positive ions. Then  phase equilibrium between  the liquid and vapor at temperature $T$ is ensured  by 
\begin{equation}
P (T,\rho_{v})=P(T,\rho_{l})  \quad \mbox{and} \quad \mu(T,\rho_{v})=\mu(T,\rho_{l}),
\end{equation}
where $\rho_{v}$ and $\rho_{l}$ are the overall particle densities in gas and liquid phases, respectively, while  $\mu= \mu_{-} + z \mu_{+}$. The pressure in each phase can still  be calculated using Eq. (\ref{pressure}). This approach  accounts for the electroneutrality of each phase,  and utilizes only one chemical potential, which significantly simplifies calculations of phase diagrams.

\subsection{Pure DH theory of charge-asymmetric lattice electrolytes}

As a  first approximation, assume that there  is no clustering between oppositely charged particles, and that  only free ions are present in the system.  The free energy density can be written  as  $\bar f=\bar f^{Id} +\bar f^{DH}$, where the first term is the entropic ideal gas contribution,  which  is given by 
\begin{equation} \label{DH.entropy}
    \bar f^{Id} = -\frac{\rho_+^*}{v_0} \ln \rho_+^* -\frac{\rho_{-}^*}{v_0} \ln \rho_{-}^* - \frac{1-\rho^*}{v_0}\ln(1- \rho^*).
\end{equation}
The subscripts  ``+'' and ``--'' denote positive and negative ions, respectively. Owing  to overall electroneutrality in the system, the densities of free ions are related to each other by 
\begin{equation}
    \rho_+^{*}=z \rho^{*}/(1+z), \;\;\; \rho_{-}^{*}=\rho^{*}/(1+z).
\end{equation}

The second term in the free energy density is the electrostatic contribution $\bar f^{DH} = \bar f_+(\rho_{+}) + \bar f_{-}(\rho_{-})$, which  is the result of  ion-ion Coulombic interactions. By solving the lattice version of the linearized Poisson-Boltzmann equation for the electrostatic potential, and  subsequently applying the  Debye charging procedure, as  was  done for  the charge-symmetric  lattice model of  electrolytes ,\cite{kobelev02} it can be easily shown that
\begin{equation} \label{f.dh.z}
    \bar f^{DH}=\frac{1}{12 v_0} \int_0^{x^2} \left[P(1)-P\left(\frac{6}{x^2+6}\right)\right]d(x^2),
\end{equation}
where
\begin{equation}
    P(\zeta)=\frac{1}{(2\pi)^{3}} \int_{-\pi}^{\pi}\int_{-\pi}^{\pi}\int_{-\pi}^{\pi} \frac{d{\mathbf k}}{1- \frac{\zeta}{3} (\cos k_1 + \cos k_2 + \cos k_3)}
\end{equation}
is the integrated lattice Green's function for the simple cubic lattice,\cite{katsura71,joyce} and $x^2=\kappa^2 a^2$ with  the reciprocal squared Debye screening length  given by
\begin{equation} \label{dlength}
    \kappa^2=\frac{4 \pi}{Dk_BT}(\rho_+q_+^2+\rho_{-}q_{-}^2)=\frac{4 \pi zq^2\rho}{D k_BT}.
\end{equation}
Now the  chemical potential for  each type of ion can be easily calculated  through Eq. (\ref{chem.potential}), which yields
\begin{equation}
    \bar \mu_i=\ln\left(\frac{\rho^*_i}{1-\rho^*}\right)+\frac{\pi }{3 T^{*}}\left[P\left(\frac{6}{x^2+6}\right)-P(1)\right],
\end{equation}
 where we  have defined the reduced temperature by 
\begin{equation} \label{temp.red}
    T^*=\frac{Dk_BTa}{z q^2},
\end{equation}
which is related to the reduced density as $ \rho^*=x^2 T^{*}/4\pi$. Then for the chemical potential of the  neutral ``molecule,'' consisting of one negative and $z$ positive ions, we obtain
\begin{equation}
    \bar \mu =(1+z)\ln\left(\frac{\rho^*}{1-\rho^*}\right) +(1+z)\frac{\pi}{3 T^*}\left[P\left(\frac{6}{x^2+6}\right)-P(1)\right]+C(z),
\end{equation}
where the  density-independent constant  is $C(z)=\left[z \ln z -(1+z) \ln(1+z) \right]$. Having the  chemical potentials and free energy, we can calculate  the pressure using Eq. (\ref{pressure}) to obtain
\begin{equation} \label{pres.dh}
    \bar p v_0 = -\ln(1-\rho^*) + \frac{1}{12}\left[x^2P\left(\frac{6}{x^2+6}\right)-\int_0^{x^2}P\left(\frac{6}{x^2+6}\right)d(x^2)\right].
\end{equation}
Note that the expression for the pressure is independent of the charge asymmetry, and the chemical potential of the ``molecule'' is $(1+z)/2$ times the value for the 1:1 electrolytes (if we neglect the constant $C(z)$  which does not affect  the thermodynamics of the system). Then, the predicted phase separation and the critical point,
\begin{equation}
    T^*_c=0.1018, \;\;\;\; \rho^*_c=0.0996,
\end{equation}
are the same as for the charge-symmetric lattice electrolytes. This result fully agrees  with the  pure DH theory in continuum space,\cite{banerjee02}  which also predicts no change in the critical parameters for charge-asymmetric electrolytes.

\section{Lattice Debye-H\"{u}ckel Theory with Clustering and Cluster-Ions Interactions  for 2:1 Electrolytes}

\subsection{Bjerrum clustering}

One of the main deficiencies of the pure DH theory, which describes the system of free positive and negative ions, is the total neglect of the ion clustering. Oppositely charged particles attract each other and form clusters in order to reduce the free energy. This process  significantly decreases  the number of free ions. In charge-symmetric lattice electrolytes  in the critical region the number of neutral ion pairs is 2-3 times larger than the number of free ions,\cite{kobelev02} and one can expect that for charge-asymmetric ionic systems  this ratio should be even larger  because of the larger free energy gain for clusters with the increase of the parameter $z$. 

The importance of ion pairing in charged particles systems was first recognized by Bjerrum.\cite{bjerrum} In his original approach  a cut-off distance between two oppositely charged ions was introduced to define a bound pair. Later, the definition of a  pair became a subject of many discussions.\cite{fisher93,levin96} However, as was shown in a careful analysis by Levin and Fisher, \cite{levin96} the precise value of this cut-off distance has little influence on the critical parameters and coexistence curves (less than 0.5\%). Meanwhile, the further problem of calculation of the electrostatic energy of the  dipole particle  poses another, new technical difficulty, since the ion pair  does not possess spherical symmetry and the problem cannot be solved exactly analytically or numerically. By approximating the ideal bispherical exclusion zone by a symmetrically centered sphere, Levin and Fisher\cite{levin96} succeeded in obtaining a precise, numerically tractable solution. For three- or four- particle clusters in charge-asymmetric continuum ionic fluids, the same strategy also yields reasonable  results, \cite{banerjee02} although the  technical complexity  of computations increases significantly.

In  lattice systems the situation is intrinsically simpler because of the discrete rather than continuous symmetry. In  $1:1$ electrolytes it allows one to define clearly an ion pair as two oppositely charged particles sitting on neighboring lattice sites.\cite{kobelev02} Similarly, for charge-asymmetric lattice electrolytes we can easily define different cluster configurations. These particles may be  viewed as independent chemical species, and the processes of clustering can be considered as a set of chemical reactions between them.

To illustrate our approach, consider a $2:1$ lattice ionic fluid in which clusters are allowed to form. Then, following Ref.\cite{banerjee02}, in addition to the free positive particles with charge $+q$ and free  negative ions with charge $-2q$, we suppose it suffices to consider  three basic types of  clusters: specifically we  include dimers carrying a charge $-q$ (with the  number density $\rho^*_2$) and two possible configurations of neutral trimers, linear and angular, with number densities $\rho^*_{3a}$ and $\rho^*_{3b}$, respectively: see Fig.1.

In the spirit of Bjerrum's  approach, we assume first that neutral clusters do not interact with  charged particles.  Then these neutral particles contribute only to the ideal free energy, and  the total  free energy density is given by
\begin{equation}
    \bar f = \bar f^{Id}(\rho_+,\rho_{-},\rho_2,\rho_{3a},\rho_{3b}) + \bar f^{El}_+ +\bar f^{El}_- + \bar f^{El}_{2}.
\end{equation}
To calculate the free energy of the charged dimers  $f^{El}_{2}$,  we adopt a view of them as a combination  of neutral symmetric $(+,-)$ dipoles, which occupy two neighboring sites, and  single-site particles (with the charge $-q$) which sit on the top of negative part of the dipoles, i.e., $(+,2-)=(+,-)+(-)$.  Then  the electrostatic free energy densities  for the free ions and charged dimers are given by the corresponding expressions from the pure DH theory, although  with a  different inverse squared Debye screening length, namely,
\begin{equation} \label{x^2.dhbj}
    \kappa^2 = \frac{4 \pi}{Dk_BT}(\rho_+q^2+4\rho_{-}q^2 + \rho_2 q^2).
\end{equation}

To determine the ideal gas (entropic) contribution to  the free energy  one  needs to know the corresponding densities and chemical potentials.  As shown in Fig.1, we identify  five different species  in the $2:1$ lattice Coulombic system. Let us  define  $\alpha$ as  a set  of free positive and negative ions,  $\beta$ as a set  of negatively charged dimers and free positive ions, $3a$ as  the linear neutral trimers, and  $3b$ as the  neutral angular trimers. Then the clustering in these system can be described by three chemical reactions
\begin{eqnarray}
    &\alpha& \stackrel{K_1}{\rightleftharpoons} \beta, \\
    &\beta&  \stackrel{K_{2a}}{\rightleftharpoons} 3a,   \\
    &\beta&  \stackrel{K_{2b}}{\rightleftharpoons} 3b,
\end{eqnarray}
with the corresponding  equilibrium  constants $K_1$,$K_{2a}$ and $K_{2b}$. The chemical equilibrium in the system is described  by the following relations between chemical potentials
\begin{equation} \label{chem.eq.2:1}
  2\mu_+ + \mu_{-}   = \mu_+ + \mu_2  = \mu_{3a} = \mu_{3b}.
\end{equation}

The chemical potential for each  particle can be computed by first  separating the entropic and electrostatic parts, i.e., $\bar\mu_i=\bar\mu^{Id}_i+\bar\mu^{El}_i$, where the latter is  determined simply  by $\bar\mu^{El}_i=-\partial \bar f^{El}/\partial \rho_i$. For the ideal-gas parts of the chemical potentials the situation is more complex. For free  single-site ions  the chemical potentials   can be easily found  applying  the  potential distribution theorem \cite{widom} or from simple entropic considerations,
\begin{eqnarray}
    \bar \mu_+^{Id}&=& \ln\rho_+^* - \ln(1-\rho_+^*-\rho_-^*-2 \rho_2^*-3\rho_{3a}^*-3\rho_{3b}^*), \\
    \bar \mu_-^{Id}&=& \ln\rho_-^* - \ln(1-\rho_+^*-\rho_-^*-2 \rho_2^*-3\rho_{3a}^*-3\rho_{3b}^*).
\end{eqnarray}
However, for dimers and  trimers there are  no similar exact expressions. Nevertheless, these entropic contributions to the chemical potentials  can be estimated by applying  the Bethe approximation,  \cite{nagle} which has been successful  in the DH  theory of $1:1$ lattice electrolytes. \cite{kobelev02} Note, that in lattice electrolyte systems phase transitions predominantly take place  in the low-density regimes, and thus the error in  using the Bethe approximation is expected to be  small.

Defining  the activities of the particles  via $z_{i}$, for every species  we have
\begin{equation} \label{activ.2:1}
    z_i= \zeta_i/\Lambda_i^{3 n} e^{\mu_i},
\end{equation}
where $n$ is the number of sites occupied by the particle, $\zeta_i$ is the corresponding internal partition function and $\Lambda_i$ is the de Broglie wavelength (see Ref.\cite{levin96}). For the latter the equality holds $\Lambda_+=\Lambda_{-}=\Lambda_2=\Lambda_{3a}=\Lambda_{3b}$. We refer to the Appendix A for a detailed calculation of the dimers and trimers activity using the Bethe approximation method, and we give here only the final expressions for activities,
\begin{eqnarray}
     z_2 &=& \frac{\rho_2^* (1-\frac{1}{3}\rho_2^*)}{(1-\rho_+^*-\rho_-^*-2 \rho_2^*-3\rho_{3a}^*-3\rho_{3b}^*)^2}e^{\bar \mu_2^{El}},  \label{z_2}\\
     z_{3a} &=& \frac{\rho_{3a}^* (1-\frac{1}{2}\rho_2^*)^2}{(1-\rho_+^*-\rho_-^*-2 \rho_2^*-3\rho_{3a}^*-3\rho_{3b}^*)^3}e^{-1/(2-\rho_{3a})}e^{\bar \mu_{3a}^{El}}, \label{z_3a}\\
     z_{3b} &=& \frac{\rho_{3b}^* (1-\frac{1}{2}\rho_2^*)^2}{64(1-\rho_+^*-\rho_-^*-2 \rho_2^*-3\rho_{3a}^*-3\rho_{3b}^*)^3}e^{-1/(2-\rho_{3b})}e^{\bar \mu_{3b}^{El}}. \label{z_3b}
\end{eqnarray}
Once the activities are known, the chemical potentials are derived by utilizing Eq.(\ref{activ.2:1}) for each species, and the free energy is obtained by integrating the chemical potentials. Finally, we arrive at the chemical potential of an $\alpha$-``molecule'' (which consists of two positive and one negative ions) in the form
\begin{equation} \label{mu.a.bj}
    \bar \mu_\alpha^{DHBj} =  \ln\rho_+^* + 2\ln\rho^*_--3\ln(1- \rho^*_+  -  \rho^*_--2\rho^*_2 - 3\rho^*_{3a} - 3\rho^*_{3b}) + \bar\mu^{El},
\end{equation}
and the reduced pressure
\begin{eqnarray} \label{pressure.bj} \nonumber
    \bar p^{DHBj}v_0 &=&  -\ln(1- \rho^*_+  -  \rho^*_--2\rho^*_2 - 3\rho^*_{3a} - 3\rho^*_{3b})+3\ln(1-\rho^*_2/3)+3\ln(1-\rho^*_{3a}/2) \\
         &+&3\ln(1-\rho^*_{3b}/2)-\frac{\rho_{3a}^*}{2-\rho^*_{3a}}-\frac{\rho_{3b}^*}{2-\rho^*_{3b}}+\bar p^{El} v_0,
\end{eqnarray}
The electrostatic part of the chemical potential and the pressure have the same form as in the pure DH theory (\ref{pres.dh}), namely,
\begin{eqnarray}
    \bar \mu^{El} &=& \frac{\pi}{3 T^*}\left[P\left(\frac{6}{x^2+6}\right)-P(1)\right], \\
     \bar p^{El} &=& \frac{1}{12}\left[x^2P\left(\frac{6}{x^2+6}\right)-\int_0^{x^2}P\left(\frac{6}{x^2+6}\right)d(x^2)\right],
\end{eqnarray}
where $x^2=\kappa^2 a^2$ with $\kappa^2$ given by (\ref{x^2.dhbj}).

In terms of activities, the chemical equilibrium conditions (\ref{chem.eq.2:1}) can be presented in the form of laws of mass action 
\begin{eqnarray}
   z_+^2 z_-  K_1 &=& z_2 z_+ ,  \label{mass.2:1.1}  \\
    z_2 z_+   K_{2a}&=&  z_{3a}, \label{mass.2:1.2}   \\
    z_2 z_+   K_{2b} &=&  z_{3b}, \label{mass.2:1.3}
\end{eqnarray}
As can be seen from the chemical equilibrium conditions (\ref{chem.eq.2:1}) and the definition of activities (\ref{activ.2:1}),  the association constants in (\ref{mass.2:1.1})-(\ref{mass.2:1.3}) are equal to the internal partition functions of dimers and trimers,
\begin{equation}
    K_1 = \zeta_1(T),\;\;\; K_{1} K_{2a} = \zeta_{3a}(T), \;\;\; K_{1} K_{2b}=\zeta_{3b}(T).
\end{equation}
The above definitions of the association constants lead to the expressions
\begin{equation}\label{const.i}
    K_i = v_0 c_i \sum_{j=1}^n e^{-u_{ji}/k_BT}
\end{equation}
where $u_{ji}=q_j\varphi_j$ is the electrostatic energy of $j$th ion  of a cluster particle  $i$ in the potential $\varphi_j$, which is  due to the other ions entering the multimer cluster. The coefficient  $c_i$ in Eq.(\ref{const.i}) is an entropic factor which takes into account all allowable configurations of the cluster. The electrostatic potentials $\varphi_j$  can be determined  by solving the lattice version of the linearized Poisson-Boltzmann equation
\begin{equation}
    \Delta \varphi_j = -\frac{4\pi}{D v_0} \sum_{k\ne j}q_k \delta({\mathbf r}_k),
\end{equation}
in which $q_k$ and ${\mathbf r}_k$ are the charge and the position of the $k$th ion entering the  multimer cluster. Since lattice Coulomb potentials can be calculated  exactly in numerical terms,\cite{kobelev02} we obtain the following expressions for association constants
\begin{eqnarray}
    K_1 &=& 6 v_0 \exp\left[\frac{1.08152}{T^*}\right], \label{K_1}\\
    K_{2a} &=&  v_0 \exp\left[\frac{0.812033}{T^*}\right], \label{K_2a}\\
    K_{2b} &=& 4 v_0 \exp\left[\frac{0.734737}{T^*}\right]. \label{K_2b}
\end{eqnarray}

Substituting (\ref{K_1})-(\ref{K_2b}) and the expressions for activities (\ref{z_2})-(\ref{z_3b}) into the expressions for chemical equilibrium  (\ref{mass.2:1.1})-(\ref{mass.2:1.3}) yields a set of  equations, which define implicitly the dimer and trimer densities $\rho_2^*$, $\rho_{3a}^*$ and $\rho_{3b}^*$ in terms  of the monomer densities $\rho^{*}_+$ and $\rho^{*}_-$. The electroneutrality of the system requires that $\rho^{*}_{+}=2 \rho^{*}_{-} + \rho^{*}_{2}$. Then the chemical potential and the pressure can be expressed in terms of only one variable, the total reduced density $\rho^{*}=\rho^*_{+} + \rho^{*}_{-}  + 2\rho^*_2  + 3\rho^*_{3a} + 3\rho^*_{3b}$, which allows for  the construction of the coexistence curve: see Fig.2.

As in the case of charge-symmetric  continuum \cite{levin96} and lattice\cite{kobelev02} electrolytes, the coexistence curve in the DHBj approximation  has an unphysical banana-like shape. This is related to the fact that,  as the temperature is lowered, the number of neutral clusters quickly grows, and this  depletes the number of free charges present in the system. Since it is the density of free charges that plays the role of the  order parameter  and  governs the gas-liquid transition, the phase separation takes place at higher overall densities. At this level of approximation,  neutral clusters are electrically inactive, and hence contribute only to the hard-core part of the free energy, which is the same for both phases. Thus, DHBj theory simply superimposes the pressure of ideal gas of clusters on the pure  DH pressure, and both sides of the coexistence curve shift to higher densities by equal amounts. The critical density is now  substantially higher, $\rho^{*}_{c} \approx 0.0807$, while the critical temperature   slightly decreases, $T^{*}_{c} \approx 0.099.$

\subsection{Cluster-Ion interactions}

The predictions of DHBj theory for lattice ionic systems  are thermodynamically unreasonable and should be corrected by taking into account the effects of interactions between multimers in clusters and free ions.\cite{levin96,kobelev02} As shown earlier for continuum and lattice electrolytes,\cite{levin96,kobelev02} these  solvation effects eliminate the unphysical banana-shape phase coexistence curves. It is reasonable to expect that these cluster-ion interactions are even more important in charge-asymmetric ionic systems due to a larger fraction of neutral clusters in equilibrium with free ions.

The exact calculations of interactions between the multimers and free ions are very complicated. Instead   we use a reasonable assumption to obtain closed analytic expressions. We approximate the neutral clusters as a set of overlapping non-interacting symmetric dipoles. For example, the neutral trimers  can be viewed as a combination of two dipoles, i.e., $(+,2-,+)=(+,-)+(-,+)$. The charged clusters, as we discussed above, are approximated as symmetric dipoles overlapping with free ions, i.e., $(+,2-)=(+,-)+(-)$. We expect that differences  between our approximation and the  exact cluster-ion interaction  contribution to the free energy  to be small since the corrections correspond to higher  order dipole-dipole interactions, and thus can be neglected.

According to our approximation the reduced density of dipoles from all cluster particles  in the system is equal $\rho^*_2 +2\rho^*_{3a}+2\rho^*_{3b}$. Since the energy of solvation of a single free dipole in the lattice electrolyte system  is known exactly,\cite{kobelev02} the cluster-ion interactions contribute to the free energy density as
\begin{equation}
  \bar f^{CI}  =(\rho^*_2 +2\rho^*_{3a}+2\rho^*_{3b})\frac{\pi q^2 a^2}{21 D k_{B}T v_0^{2}}\left[-\frac{x^2}{2}+\frac{1}{x^2}\int_0^{x^2}G(x^2)d(x^2)\right],
\end{equation}
where
\begin{equation}
    G(x^2)=\frac{x^2(x^2+7)}{x^2+6}P\left(\frac{6}{x^2+6}\right).
\end{equation}
Then  the corresponding contributions to the  chemical potentials and pressure are given by
\begin{equation}
    \bar \mu_\alpha^{CI} = \frac{2\pi^2 (\rho^*_2+2\rho^*_{3a}+2\rho^*_{3b})}{21 (T^*)^2} \left[\frac{1}{2} - \frac{1}{x^2}G(x^2) + \frac{1}{x^4}\int_0^{x^2}G(x^2)d(x^2)\right]
\end{equation}
and
\begin{equation}
    \bar p^{CI} v_0 = \bar \mu_\alpha^{CI} x^2T^*/4\pi,
\end{equation}
which must be added to the values provided by Eqns. (\ref{mu.a.bj}) and (\ref{pressure.bj}) in order  to obtain the complete  expressions for the  chemical potential and  pressure.

The full DHBjCI theory predicts a phase separation as exhibited  in Fig.2. As in the  charge-symmetric lattice and continuum electrolytes,\cite{levin96,kobelev02} taking into account the solvation of  clusters by the residual free ions eliminates  the unphysical  banana-like shape of the coexistence curve.  The critical parameters are now  given by
\begin{equation}
    T^*_c = 0.08735,\;\;\; \rho^*_c = 0.05001,\;\;\; Z_c\equiv p_{c}/\rho_{c} k_BT = 0.2431.
\end{equation}
Comparison with the critical parameters of 1:1 lattice ionic system indicates that  the critical temperature is about 11$\%$ lower, while the critical density is 1.67 times higher. These trends, the decrease in the critical temperature and   inrease in the critical density, are in  agreement with the results of continuum calculations for charge-asymmetric electrolytes.\cite{banerjee02}

It is interesting to estimate the relative amounts of different species in the critical region. In terms of molar fractions $y_i=n_{i} \rho_i^*/\rho^*$ ($n_{i}$ is the size of the particle), our theoretical  approach  predicts
\begin{eqnarray}
    y_+&=&0.1002, \;\;\; y_- = 0.0388, \;\;\; y_2 = 0.0452, \\ \nonumber
    y_{3a}&=& 0.0075, \;\;\; {\rm and} \;\;\; y_{3b}=0.8083.
\end{eqnarray}
As expected, the fraction  of neutral trimers significantly exceeds  the fraction  of free charges. Also, the number of charged dimers is very low, which can be attributed to their propensity  to combine with free ions to form  energetically more favorable neutral trimers. What is surprising, at first glance,  is that the ratio of linear trimers to angular trimers is much less than unity. Indeed, because electrostatic interactions favor a linear arrangement of the ions in a trimer cluster, these clusters  are more energetically stable and should prevail over angular trimers. However, these naive arguments do not take into account  the entropic considerations.  First of all, the angular trimers have more different possible configurations  than the linear trimers. In addition, they are  more compact and  can be packed more densely. As a consequence, there are  more ways to arrange angular trimers on the lattice and they dominate trimer clusters.

\section{Lattice DHBjCI Theory for 3:1 and 4:1  Electrolytes}

The thermodynamic calculations for $z=3$ asymmetric lattice electrolytes can be performed following the method presented  in full  detail in  Sec.III. As shown in Fig.3, there are seven basic clusters to be considered in this system, namely, single-site  positive and negative ions, dimers with the charge $-2q$, linear and angular trimers (with the charge $-q$), and two types of neutral tetramers, which are connected by the network of six chemical reactions. Note that we do not consider  clusters  with  bonds between the same charges. For example, we assume that  charged trimers  have the configurations (+,3$-$,+), but not (+,+,3$-$). Indeed, these particles are less stable and their contributions to free energies can be neglected. 

In addition, as in  the case of 2:1 lattice Coulombic systems, we view the clusters  as combinations of noninteracting (+,$-$) symmetric dipoles  and single-site charges. The activities of multimers are calculated by applying the Bethe approximation method outlined in the Appendix. These approximations  allow us to calculate the free energy of the system as a sum of the corresponding entropic, electrostatic and cluster solvation terms for each particle. The resulting phase coexistence curve is given in Fig.4. The critical parameters are
\begin{equation}
T^{*}_{c}=0.0688, \quad \quad \rho^{*}_{c}=0.0847.
\end{equation} 
The molar fractions of different clusters  in the critical region are
\begin{eqnarray}
    y_+&=&0.08872, \;\;\; y_- = 0.00347, \;\;\; y_2 = 0.0088,  \\ \nonumber
    y_{3a}&=& 0.0033, \;\;\;   y_{3b}=0.2049, \;\;\;  y_{4a}=0.3043,  \;\;\; y_{4b}=0.3864.
\end{eqnarray}
As expected, neutral clusters again dominate. 
  
Analogous thermodynamic calculations  can be done for 4:1 lattice electrolytes. However, the number of different types of clusters and the number of chemical reactions between them become fairly  large, and the full thermodynamic analysis is difficult to complete. Instead, we  focus on the critical region of the system where calculations can be completed. The resulting critical parameters are 
\begin{equation}
T^{*}_{c}=0.060, \quad \quad \rho^{*}_{c}=0.148.
\end{equation} 
The relative density  of all neutral clusters here  is about 87$\%$. 

Full phase diagrams for $z:1$ lattice ionic systems ($z=1,2$ or 3) calculated using DHBjCI approach are exhibited in Fig.4.

\section{Discussion}

Our analysis of charge-asymmetric  lattice electrolytes using  Debye-H\"{u}ckel theory with Bjerrum clustering and cluster-ion interactions indicates that charge asymmetry strongly influences the thermodynamics, especially in the critical region. It is found that critical temperatures  decrease, while critical densities  increase with charge asymmetry. Our theoretical predictions for critical parameters are compared in Fig.5  with the results of  continuum  Monte Carlo simulations\cite{camp99,pana02,yan02} and with the predictions of the continuum DHBjCI theory. \cite{banerjee02} 

Clearly, the overall consistency between the lattice and continuum DHBjCI  approaches and computer simulations results indicates that the Debye-H\"{u}ckel method correctly captures the physics of phase transitions in these ionic systems.  In  Coulomb systems with charge asymmetry the formation of clusters is strongly favored. Clusters contribute to electrostatic interactions much less than free ions, and the effective electrostatic energy per particle decreases. However, the phase separation is driven by charged particles, and the temperature is normalized by the strongest electrostatic interaction between the positive and negative ions: see Eq.(\ref{temp.red}). Therefore, the reduced critical temperature falls. 

Cluster formation  also significantly  depletes  the number of free ions. At  the critical point, the molar fractions of free single-site positive and negative ions are 21.0$\%$, 13.8$\%$, 9.2$\%$ and 4.2$\%$ for 1:1, 2:1, 3:1 and 4:1 lattice electrolyte systems, respectively.   As a consequence, larger overall densities are required to  achieve the  phase separation, and the reduced critical density increases.

At the same time, the relative numbers of neutral particles as a function of charge asymmetry shows a non-monotonic  behavior. The molar fractions of neutral clusters are  equal to 79\% in 1:1 ionic system,  81\% in 2:1 electrolytes, 69.1\% in 3:1 ionic system  and about 88\% in 4:1 electrolytes. The decrease of the molar fraction of neutral clusters in the critical region  in   3:1 system as compared to  2:1 electrolytes  has also been found in continuum calculations.\cite{banerjee02}  The actual number of neutral clusters depends on two factors: the free energy gain of cluster formation and  the temperature (through the  equilibrium constant). It is possible that in 3:1 electrolytes the lowering of the critical temperature is not enough to overcome the repulsion between the positive ions in neutral tetramers, while in 4:1 ionic systems the decrease in critical temperature is  enough to favor more  strongly  the formation of neutral clusters.

The absolute values of the critical temperature in our charge-asymmetric lattice models is larger than the Monte Carlo simulation  data\cite{camp99,pana02} by a factor of $1.68 - 1.94$, with lesser  factors corresponding to higher $z$.  This agrees with the general feature of lattice models to over-estimate the  values of critical temperatures in comparison with  continuum models.   Note, however,  that the simulations are performed with the continuum potential, while our theory employs  the  lattice Coulombic potentials. In addition, the reduced temperature (\ref{temp.red})  is  normalized by the energy of two ions interaction via {\it continuum} or $1/r$ Coulombic potential. However, the  lattice potential differs from $1/r$ at short distances,\cite{kobelev02}  and normalizing the temperature by the energy of lattice Coulombic interactions of nearest neighbors would lower the  temperatures  by an additional 8$\%$.

The absolute values of the critical densities in our theoretical approach are only  38-69$\%$ of the current Monte Carlo estimates.\cite{camp99,pana02} This fact  reflects  the discrete nature of the lattices  and  that not all possible clusters have been taken  into account.

\section{Conclusions}

We have investigated the effects of charge asymmetry on thermodynamics and critical properties of the lattice ionic systems using Debye-H\"{u}ckel theory with Bjerrum clustering and cluster-ion interactions. Phase diagrams for $z:1$ lattice electrolytes have been  obtained for $z=2$ and 3, while for $z=4$ the location of the critical point was determined. Our results agree well with the Monte Carlo simulations and with the continuum  Debye-H\"{u}ckel theory,\cite{banerjee02} predicting that the increase in the charge asymmetry  lowers the critical temperature and increases the critical density.

Our theoretical approach may be extended in several directions. In this paper we investigated charge asymmetric electrolyte systems on the simple cubic lattices. It is interesting to consider the lattices with different symmetry such as body-centered cubic and face-centered cubic lattices, for which thermodynamic calculations  for charge symmetric systems  are already performed.\cite{kobelev02} Another interesting question is how the charge asymmetry will affect the ionic systems in two dimensions, where interesting phase transitions  may appear.

\section*{Acknowledgments}

Acknowledgment is made to the Donors of the American Chemical Society Petroleum Research Fund (Grant No. 37867-G6) for support of this research. The authors also acknowledge the support of the Camille and Henry Dreyfus New Faculty Awards Program (Grant No. NF-00-056). The authors would like to thank Professor M.E. Fisher for critical comments and useful suggestions, and for providing us with a draft manuscript of his work.

\newpage

\appendix

\section{The Bethe approximation for multimers}

Consider a system  of $N_m$ multimers  (each particle occupies $m$ lattice sites), which interact with each other only via on-site exclusion, on a simple cubic lattice with the total number of sites $N$. We can calculate the entropy of the multimer distribution following a direct method of counting probabilities in the Bethe approximation  employed by Nagle\cite{nagle}. (For other methods of calculating the partition function for lattice systems see  Refs.\cite{kikuchi, burley}.) The Bethe approximation does not take into account any correlations around cycles and thus it is exact for lattices with only tree-like  paths  (Bethe lattices). However for $3D$ cubic lattices this approximation  gives fairly good results.

Consider, first,  a system consisting only  of dimers. Then  the number of ways to arrange $N_2=N\rho^*_2$ dimers on the simple cubic lattice can be approximated by\cite{nagle}
\begin{equation}
    W_2(\rho^*_2)=\left[\left(\begin{array}{c} N \\ 2 N\rho^*_2 \end{array}\right)6^{2N\rho^*_2}\right]\left[\left(\frac{2\rho^*_2}{6}\right)^{1/2}\left(1-\frac{2\rho^*_2}{6}\right)^{5/2}\right]^{2N\rho^*_2}\left[\left(1-\frac{2\rho^*_2}{6}\right)^{6/2}\right]^{N(1-2\rho^*_2)}.
\end{equation}
Here the first square bracket represents the total number of ways to arrange $N\rho^*_2$ dimer vertices on the lattice, taking into account six possible orientations for each dimer, provided one its vertex is
fixed. The second bracket gives the probability that a dimer vertex configuration is compatible with all its nearest-neighbors' vertex configurations. Here $2\rho^*_2/6$ is the probability of finding  dimer's edge
along certain lattice direction starting from a given point,  and $(1-2\rho^*_2/6)^5$ ensures there is only one dimer at this point. The third square bracket factor is the probability that no dimers occupy
$N(1-2\rho^*_2)$ empty lattice sites. The square roots are taken in order not to compute   the probabilities  twice. The Bethe approximation for dimer's activity on  the sc lattice then yields
\begin{equation}
    z  = \exp\left(-\frac{1}{N}\frac{d }{d \rho^*_2}\ln W_2(\rho^*_2)\right) = \frac{\rho^*_2/3  (1-\rho^*_2/3)}{(1-2\rho^*_2)^2}.
\end{equation}

Similar calculations can be performed for the systems containing  trimers, tetramers or higher order multimers.  However, since now not all multimer vertices are equivalent, we have to account for each type of vertex separately. 

In particular, consider the system with  $N_{3}=  N\rho^*_3$ trimers. Then there are  $3 N\rho^*_3$ trimer vertices, with $N\rho^*_3$ ``center'' and  $2 N\rho^*_3$  ``end'' vertices. For linear trimers the number of possible orientations is 3 if we fix a ``center'' vertex and 6 if we fix an ``end'' vertex. Then the number of ways to arrange linear trimers on the lattice is given by
\begin{eqnarray} \nonumber
    W_3(\rho^*_3)&=&\left[\left(\begin{array}{c} N \\  3N\rho^*_3 \end{array}\right)6^{2N\rho^*_3}3^{N\rho^*_3}\right]\left[\left(\frac{3\rho^*_3}{6}\right)^{1/2}\left(1-\frac{3\rho^*_3}{6}\right)^{5/2}\right]^{2N\rho^*_3} \\
    &\times &\left[\left(\frac{3\rho^*_3}{6}\right)^{2/2}\left(1-\frac{3\rho^*_3}{6}\right)^{4/2}\right]^{N\rho^*_3}\left[\left(1-\frac{3\rho^*_3}{6}\right)^{6/2}\right]^{N(1-3\rho^*_3)}.
\end{eqnarray}
Again the first square bracket gives the maximum possible number of distinct trimer vertices configurations, the second square bracket accounts for  the probability that all ``end'' vertex configurations are compatible with their nearest-neighbors' configurations,  the third bracket corresponds to the ``center'' vertices' compatibility, and the last factor ensures that there are  $N(1-3\rho^*_3)$ lattice  sites free
of trimers. Then the activity of linear trimers is given by
\begin{equation}
    z_{3a}=\frac{\rho^*_{3a} (1-3\rho^*_{3a}/6)^2}{(1-3\rho^*_{3a})^3}e^{-1/(2-\rho^*_{3a})}.
\end{equation}
For angular trimers, the only difference is that now there are 24 distinct trimer orientations if we start from the  ``end'' vertices,  and  12 orientations if we start from the ``center'' vertex. This yields
\begin{equation}
    z_{3b}=\frac{1}{64} \frac{\rho^*_{3b} (1-3\rho^*_{3b}/6)^2}{(1-3\rho^*_{3b})^3}e^{-1/(2-\rho^*_{3b})}.
\end{equation}
Similarly,  the activity of  tetramers  is given by the following expression
\begin{equation}
    z_4 = C \frac{\rho^*_{4} (1-2\rho^*_4/3)^2}{(1-4\rho^*_{4})^4}e^{-1/(1-2\rho^*_{4}/3)},
\end{equation}
where the coefficient $C$ equals to $1/96$ for planar tetramers and $C=1/128$ for non-planar tetramers.

\begin{figure}[h]
\begin{center}
\vskip 1.5in
\unitlength 1in
\begin{picture}(3.4,2.8)
\resizebox{3.4in}{2.8in}{\includegraphics{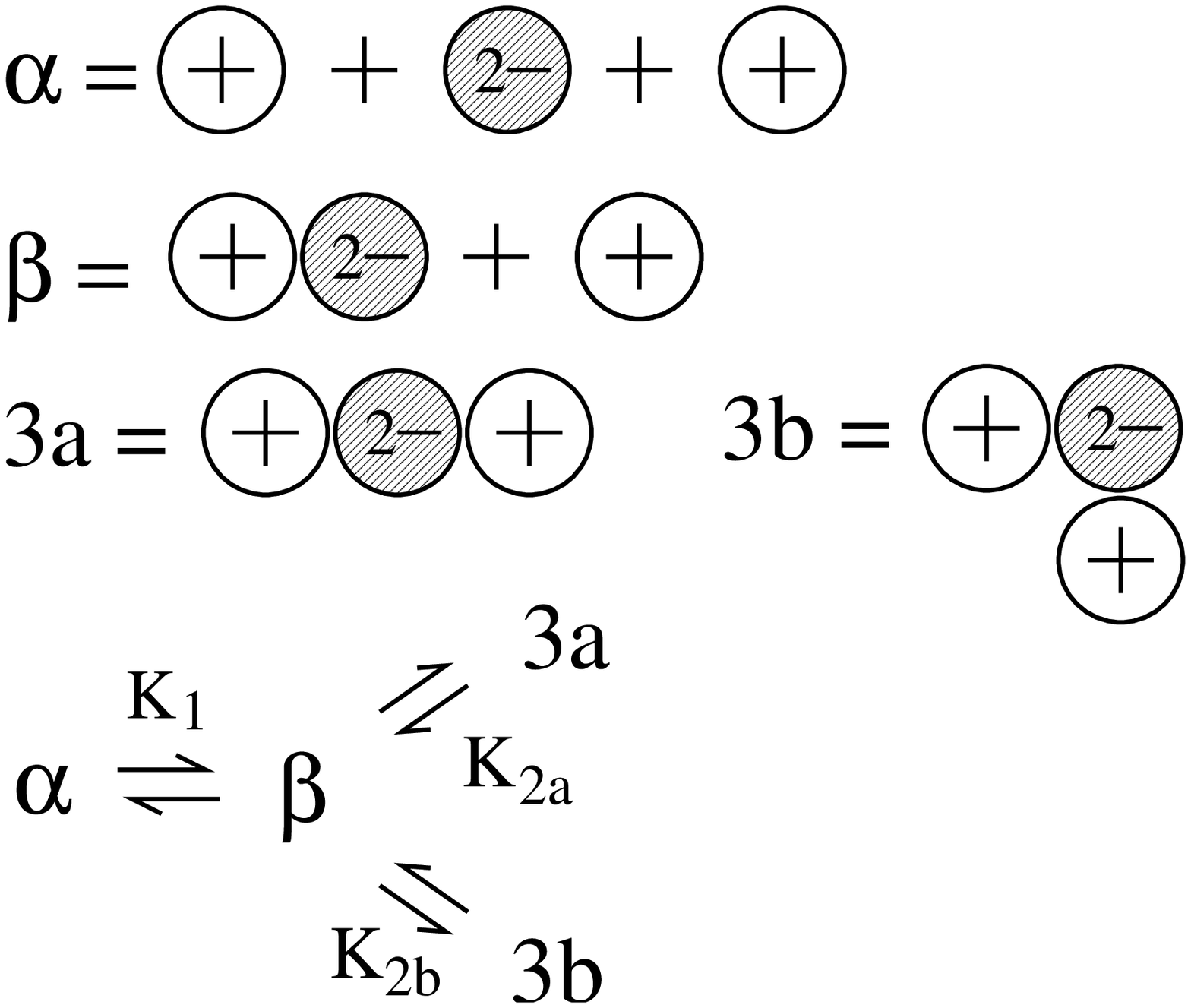}}
\end{picture}
\vskip 3in
 \begin{Large} Fig.1 \end{Large}
\end{center}
\vskip 3in
\end{figure}

\begin{figure}[h]
\begin{center}
\vskip 1.5in
\unitlength 1in
\begin{picture}(4.5,3.5)
\resizebox{4.5in}{3.5in}{\includegraphics{fig2.eps}}
\end{picture}
\vskip 3in
 \begin{Large} Fig.2 \end{Large}
\end{center}
\vskip 3in
\end{figure}

\begin{figure}[h]
\begin{center}
\vskip 1.5in
\unitlength 1in
\begin{picture}(3.5,5.2)
\resizebox{3.5in}{5.2in}{\includegraphics{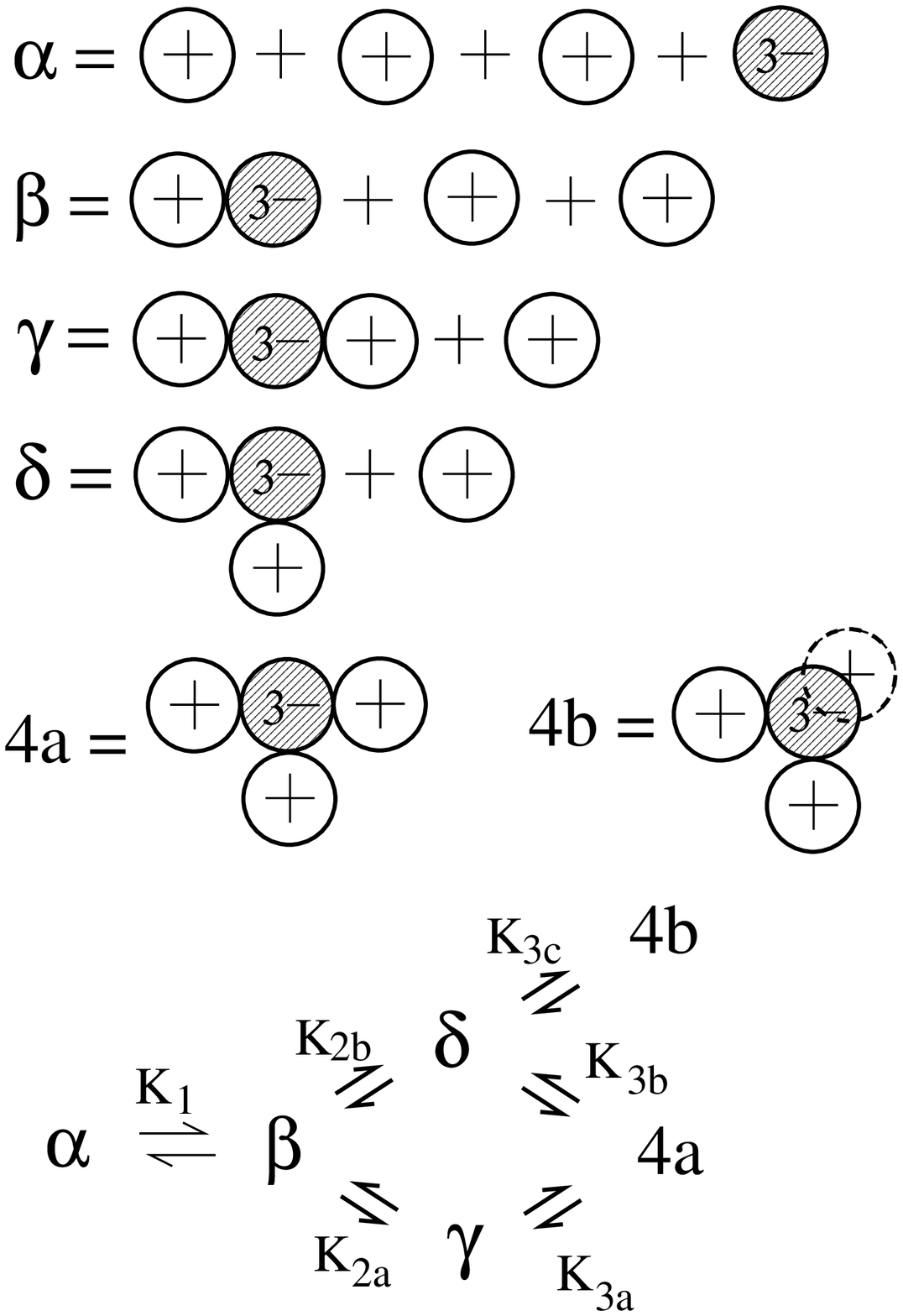}}
\end{picture}
\vskip 3in
 \begin{Large} Fig.3 \end{Large}
\end{center}
\vskip 3in
\end{figure}

\begin{figure}[h]
\begin{center}
\vskip 1.5in
\unitlength 1in
\begin{picture}(4.5,3.5)
\resizebox{4.5in}{3.3in}{\includegraphics{fig4.eps}}
\end{picture}
\vskip 3in
 \begin{Large} Fig.4 \end{Large}
\end{center}
\vskip 3in
\end{figure}

\begin{figure}[h]
\begin{center}
\vskip 1.5in
\unitlength 1in
\begin{picture}(4.5,3.5)
\resizebox{4.5in}{3.3in}{\includegraphics{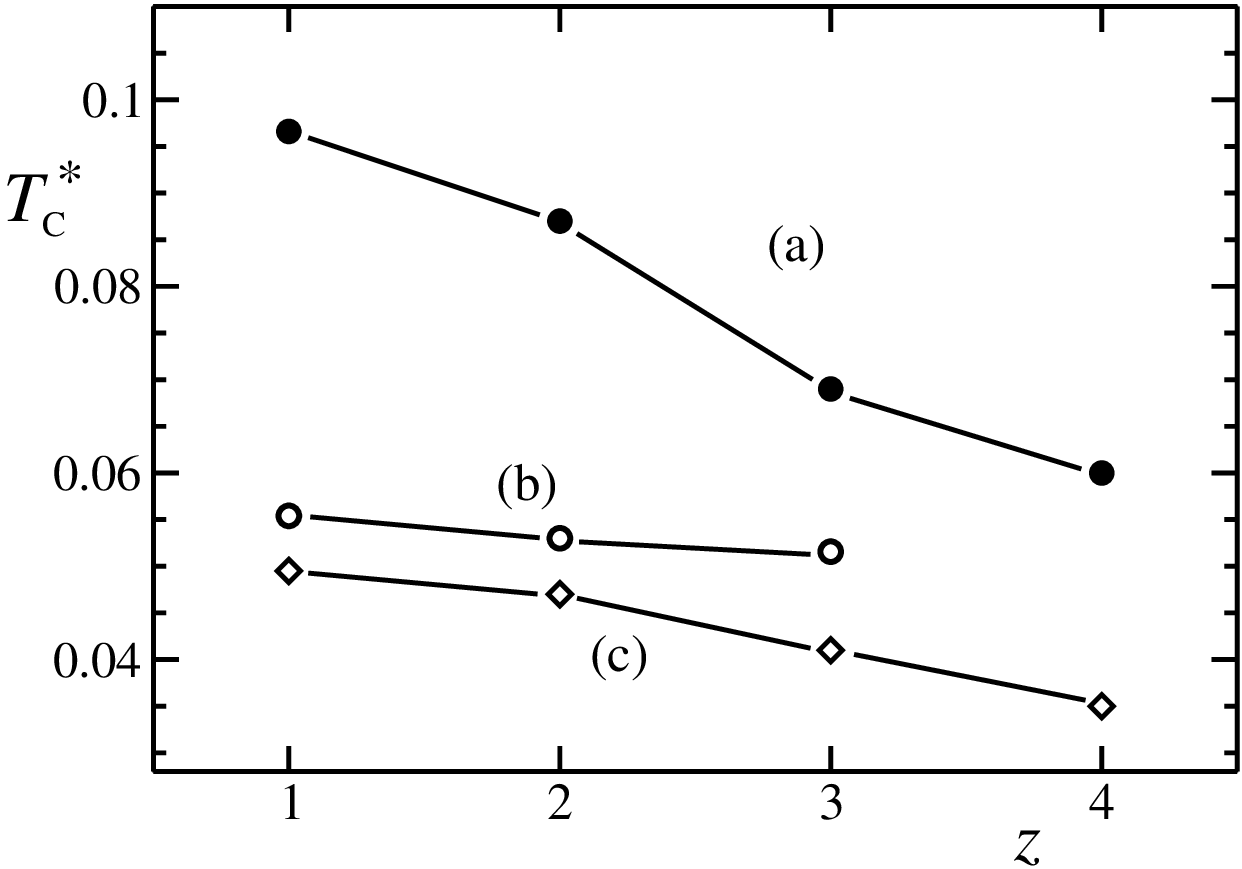}}
\end{picture}
\vskip 3in
 \begin{Large} Fig.5A \end{Large}
\end{center}
\vskip 3in
\end{figure}

\begin{figure}[h]
\begin{center}
\vskip 1.5in
\unitlength 1in
\begin{picture}(4.5,3.5)
\resizebox{4.5in}{3.5in}{\includegraphics{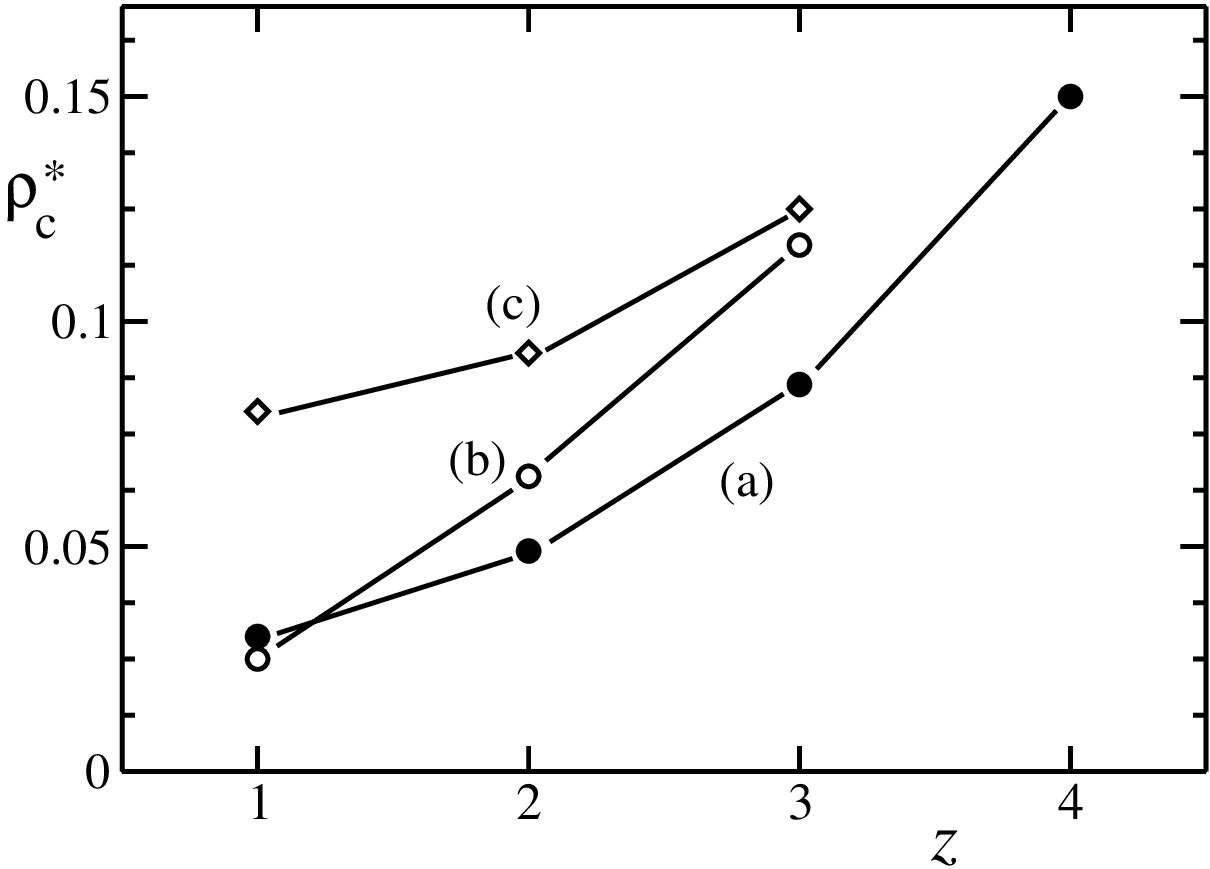}}
\end{picture}
\vskip 3in
 \begin{Large} Fig.5B \end{Large}
\end{center}
\vskip 3in
\end{figure}

\newpage

Figure captions. \\\\

Fig.1.  Different charged and neutral particles and chemical reactions in 2:1 lattice electrolyte systems.\\

Fig.2.  Phase diagrams for 2:1 lattice electrolyte: (a) pure DH theory, (b) DHBj approximation, (c)  DHBjCI theory.\\

Fig.3.   Different charged and neutral particles and  chemical reactions in 3:1 lattice electrolyte. \\

Fig.4.  Phase diagrams of  (a) 1:1, (b) 2:1, (c) 3:1 lattice electrolytes in  DHBjCI theory.\\

Fig.5.  A) Critical temperatures $T^*_c$ and B) critical densities $\rho_c^*$  as functions of charge asymmetry: (a) our lattice DHBjCI predictions, (b) continuum DHBjCI predictions,\cite{banerjee02} (c) Monte Carlo simulations.\cite{camp99,pana02,yan02}

\end{document}